\documentclass[11pt,twoside]{article}
\usepackage{graphicx,epsfig,natbib,epstopdf}
\usepackage{CS18}
\begin{document}
%
% NTA: Ignore the setcounter comment here
%\setcounter{page}{7}
%
% NTA: Update your full title here
%
\title{V4046 Sgr: Touchstone to Investigate Spectral Type Discrepancies for Pre-main Sequence Stars}
%
% NTA: enter your author name, affiliation/address information here
%
\author{Joel H. Kastner$^{1}$, Valerie Rapson$^{1}$, Benjamin Sargent$^{1}$, C.T. Smith$^{1,2}$, John Rayner$^{3}$}
\affil{$^1$Center for Imaging Science, School of Physics \& Astronomy, and Laboratory for Multiwavelength Astrophysics, Rochester Institute of Technology, 54 Lomb Memorial Drive, Rochester, NY 14623, USA; jhk@cis.rit.edu}
\affil{$^2$University of Arizona, Tucson, AZ}
\affil{$^3$University of Hawaii and NASA Infrared Telescope Facility, Honolulu, HI 96822, USA}
\begin{abstract}
%
% NTA: Update your abstract here.
%
Determinations of the fundamental properties (e.g., masses and ages) of late-type, pre-main sequence  (pre-MS) stars are complicated by the potential for significant discrepancies between the spectral types of such stars as ascertained via optical vs.\ near-infrared observations. To address this problem, we have obtained near-IR spectroscopy of the nearby, close binary T Tauri system V4046 Sgr AB with the NASA Infrared Telescope Facility (IRTF) SPEX spectrometer. The V4046 Sgr close binary (and circumbinary disk) system provides an important test case for spectral type determination thanks to the stringent observational constraints on its component stellar masses (i.e., $\sim$0.9 $M_\odot$ each) as well as on its age (12--21 Myr) and distance (73 pc). Analysis of the IRTF data indicates that the composite near-IR spectral type for V4046 Sgr AB lies in the range M0--M1, i.e., significantly later than the K5+K7 composite type previously determined from optical spectroscopy. However, the K5+K7 composite type is in better agreement with theoretical pre-MS evolutionary tracks, given the well-determined properties of V4046 Sgr AB. These results serve as a cautionary tale for studies that rely on near-infrared spectroscopy as a primary means to infer the ages and masses of pre-MS stars. 
\end{abstract}
%
%
%
%
% NTA: Here is where the body of your text goes.
%

\section{Introduction}

The potential for significant discrepancies between the spectral types of late-type pre-main sequence (pre-MS) stars as determined via visible-wavelength (hereafter ``optical'') vs.\ near-infrared observations was first recognized and documented more than 15 years ago \citep{1998ASPC..154.1709G}. Clearly, such discrepancies have important implications for studies that seek to use theoretical pre-main sequence (pre-MS) evolutionary tracks to infer the ages and masses of pre-MS stars based on spectral types --- and, hence, effective temperatures --- that have been determined via near-infrared spectroscopy. 

A case in point is the well-studied, nearby T Tauri star TW Hya. Most investigators have adopted an optical spectral type of K7Ve for TW Hya, following the original determination by \citet{herbig78}. Subsequent optical spectroscopic studies have supported or bracketed Herbig's determination; e.g., \citet{2006A&A...460..695T} found a spectral type of K6Ve, while \citet{2013ApJS..208....9P} found K8IVe. However, based on analysis of a near-infrared ($\sim$1--5 $\mu$m) spectrum, \citet{2011ApJ...732....8V} determined the spectral type of TW Hya to be much later; they found M2.5V. \citet{2011ApJ...732....8V} used this result to reassess the mass and age of TW Hya, based on comparisons with pre-MS tracks. They concluded that the (optical spectral type-based) mass and age of TW Hya had previously been overestimated, and proposed revising these estimates from $\sim$0.8 $M_\odot$ to $\sim$0.4 $M_\odot$ and from $\sim$8 Myr to $\sim$3 Myr, respectively. \citet{2011ApJ...732....8V} argued that these revised estimates were marginally consistent with existing constraints on the mass and age of TW Hya.

Thanks to the (far more) stringent constraints on its fundamental properties, the short-period binary T Tauri system V4046 Sgr AB \citep[optical spectral types K5+K7;][]{2004A&A...421.1159S,2011MNRAS.417.1747D} provides an important test case for the study of potential  discrepancies between near-IR and optical spectral type determinations. The estimated age and distance of V4046 Sgr are relatively well-established (i.e., 12--21 Myr and $\sim$73 pc, respectively) on the basis of its membership in the  $\beta$ Pic Moving Group and the presence of a comoving, early-M companion \citep{2008hsf2.book..757T,2011ApJ...740L..17K,2014MNRAS.438L..11B}. More importantly, the masses of the components of V4046 Sgr AB were precisely determined by \citet{2012ApJ...759..119R}  --- i.e., 0.90 $M_\odot$ (A) and 0.85 $M_\odot$ (B), with uncertainties of $\sim$0.05 $M_\odot$ --- from analysis of interferometric CO imaging of its extended, inclined, circumbinary disk (which yields a dynamical mass of $\sim$1.75 $M_\odot$) combined with optical radial velocity measurements of the central, short ($\sim$2.4 d) period binary.
% (which indicate nearly equal-mass components). 
\citet{2012ApJ...759..119R} further demonstated that these dynamical masses are in excellent agreement with the predictions of models describing pre-MS evolution, given photospheric temperatures corresponding to optical spectral types of K5+K7.

Hence, we set out to determine whether the apparent spectral type discrepancies just described for TW Hya might manifest themselves similarly in the case of V4046 Sgr\footnote{For a review of the V4046 Sgr binary/circumbinary disk system and a comparative summary of the properties of the TW Hya and V4046 Sgr systems, the reader is referred to \citet{kastner13} and \citet{kastner14}, respectively.} and, if so, what this might imply for near-IR-based determinations of pre-MS masses and ages more generally.

\section{Observations}

We observed V4046 Sgr with the medium-resolution SpeX spectrograph \citep{2003PASP..115..362R} on the NASA Infrared Telescope Facility (IRTF). This was the same telescope/instrument combination used by \citet{2011ApJ...732....8V} in their study of TW Hya. Data for V4046 Sgr were obtained on June 28, 2013. We obtained ten individual 120s exposures of V4046 Sgr in pairing mode using the short-wavelength cross-dispersed (SXD) mode of SpeX, and twenty 5 s exposures using the long wavelength cross-dispersed (LXD2.1) mode. The slit width was set to 0.3$''$, yielding a resolution of $\sim$2000 for the SXD spectra and $\sim$2500 for the LXD2.1 spectra. Together, these observations cover the entire 0.8 -- 5.0 $\mu$m spectral range. The airmass during these observations was 1.65-1.7, and the sky was clear, with seeing at 1.2$''$. Observations of the A0V star HD 171296 were obtained the same night, to determine the strengths of telluric absorption features \citep{2003PASP..115..389V}. Other calibration data (flat fields and arc frames) were obtained immediately after on source observations in each spectral mode. Standard data reduction was performed with Spextool v3.4 \citep{2004PASP..116..362C}. 
%(i.e., nonlinearity corrections, flat fielding, image pair subtraction, aperture definition, optimal extraction, and wavelength calibration) 
The SXD and LXD data were cleaned and median combined, and telluric features were removed from all orders. Low signal-to-noise ratio data between 1.355--1.405 $\mu$m and 1.870--1.930 $\mu$m were eliminated. The reduced SXD and LXD spectra were then merged. The resulting complete IRTF/SpeX spectrum of V4046 Sgr AB is presented in Fig.~\ref{fig:FullSpec}. 

\section{Results}

\begin{figure}[htbp]
\begin{center}
\includegraphics[width=5in]{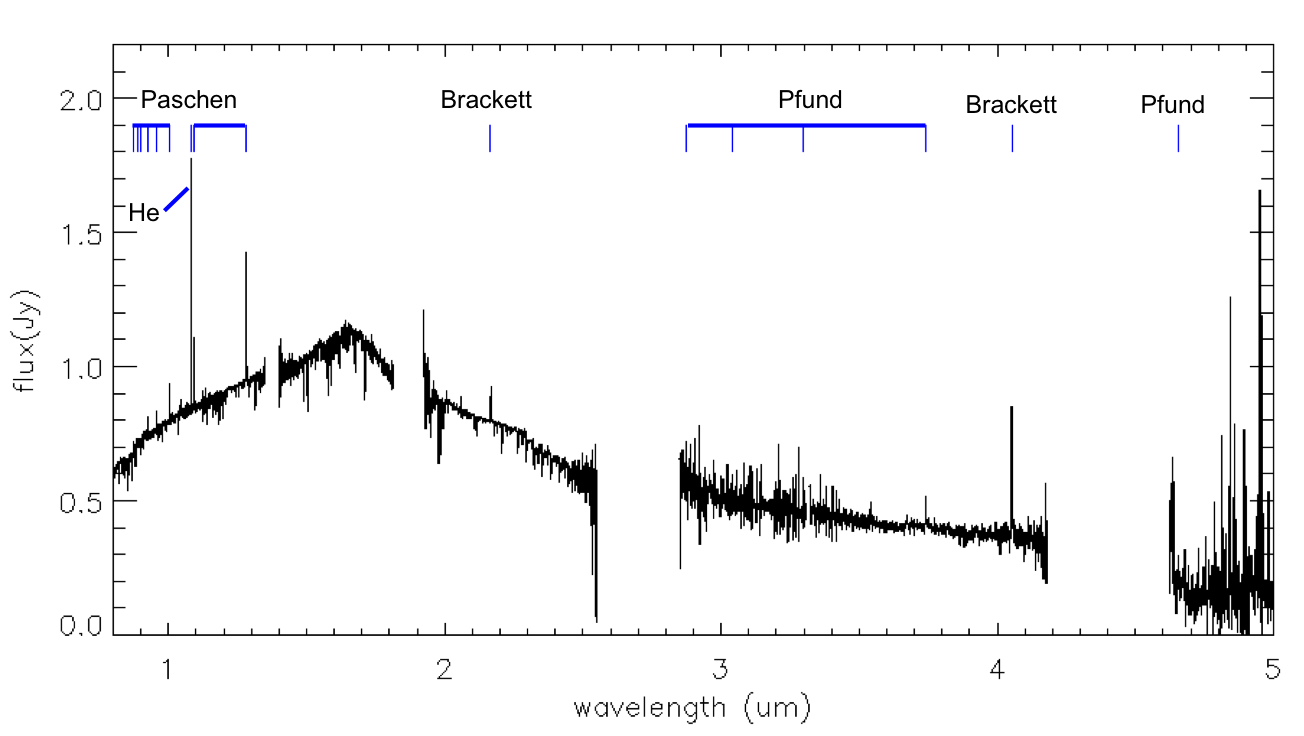}
\caption{The complete IRTF/SpeX spectrum of V4046 Sgr, with prominent emission lines indicated.
}
\label{fig:FullSpec}
\end{center}
\end{figure}

%\begin{figure}[htbp]
%\begin{center}
%\includegraphics[width=5in]{comparison_0_8-1_1um.png}
%\caption{...selected portions of the IRTF/SpeX spectrum...}
%\label{fig:SpecExamples}
%\end{center}
%\end{figure}

A relatively cursory (visual) comparison of the IRTF/SpeX infrared spectrum of V4046 Sgr AB with those of IRTF standard stars --- which are main sequence (MS) stars with optically-determined spectral types \citep{2009ApJS..185..289R} --- suggests that its near-infrared spectrum more closely resembles those of borderline late-K/early-M MS stars than that of the K5 type of its dominant component (i.e., the slightly more massive component A). More specifically, the broad, shallow absorption from TiO bandheads at $\sim$0.85 $\mu$m is indicative of a late K or early M type star, and the overall shape of the continuum from 0.8--1.8 $\mu$m and 1.9--2.6 $\mu$m resembles those of K7 and M1.5 IRTF spectral standards, respectively. As in the case of TW Hya, the hydrogen emission-line spectrum characteristic of an actively accreting T Tauri star is superimposed on this relatively red continuum (Fig.~\ref{fig:FullSpec}).

%Fig.~\ref{fig:SpecExamples} illustrates selected portions of the IRTF/SpeX spectrum of V4046 Sgr AB, with IRTF standard star spectra spanning the spectral type range K7 to M3 included for comparison. {\bf [??]}

\begin{table}
\begin{center}
%\small
%\footnotesize
\caption{\sc IR Absorption Line Equivalent Width Comparisons}% and  comparisons}
\label{tbl:EWs}
\vspace{.05in}
\begin{tabular}{lccccc}
\hline
\hline
 & & \multicolumn{3}{c}{Equivalent widths$^a$} & \\
 Species & $\lambda$ & V4046 Sgr & K5-K7 & M0-M3 & Closest matches \\
& ($\mu$m) & (\AA) & (\AA) &  (\AA) & \\ 
\hline
K {\sc i} & 1.253  &   0.44$\pm$0.01 & 0.4--0.5 & 0.4--1.0 & K5--M0.5 \\
Mg {\sc i} & 1.183  &    1.37$\pm$0.01 & 1.5--2.0 & 0.8--1.5 & M0--M1.5 \\
Mg {\sc i} &  1.504  &    5.54$\pm$0.04 & 7.0--8.0 & 2.0--6.0 & M0.5--M2 \\
Mg {\sc i} &  1.711  &    3.01$\pm$0.03 & 3.3--3.8 & 1.5--3.8 & M1--M2 \\
Na {\sc i} & 1.140  &    2.46$\pm$0.03 & 1.7--2.7 & 2.0--4.5 & K7--M2 \\
FeH & 0.990  &    1.53$\pm$0.02 & 1.0--1.8 & 2.0--5.0 & K7--M0 \\
\hline
\end{tabular}
\end{center}

%\vspace{.05in}
\footnotesize {\sc Note:} a) EWs for K5-K7 and M0-M3 spectral
standards from \citet[][their Fig.\ 3]{2011ApJ...732....8V}.\\
%b) Spectral types for V4046 Sgr and TW Hya as determined from optical
%spectroscopic analysis presented in \citet{2004A&A...421.1159S} and \citet{herbig78}, respectively.
\end{table}
To make the comparison between V4046 Sgr AB and IRTF spectral type standards more quantitative, we adopted the IDL-based tools\footnote{See https://github.com/awmann/metal.} described in \citet{2013AJ....145...52M} to perform measurements of the equivalent widths (EWs) of selected absorption features in the IRTF/SpeX spectrum of V4046 Sgr AB that are particularly sensitive diagnostics of spectral type \citep{2009ApJS..185..289R}. Selected EW measurements are presented in Table~\ref{tbl:EWs}, alongside those obtained from the IRTF/SpeX spectra of late K and early M standard stars by \citet{2011ApJ...732....8V}. It is readily apparent from these comparisons that V4046 Sgr AB, like TW Hya, displays a near-infrared spectral type significantly later than the spectral type combination determined via optical spectroscopy. Specifically, based on the aforementioned visual inspection and the EW results in Table~\ref{tbl:EWs}, we judge that the IRTF/SpeX-determined composite spectral type of V4046 Sgr AB lies in the range M0--M1 --- i.e., 3--5 spectral subclasses later than the optically-determined composite spectal type of K5+K7.

\section{Discussion}

\begin{figure}[htbp]
\begin{center}
\includegraphics[width=5in]{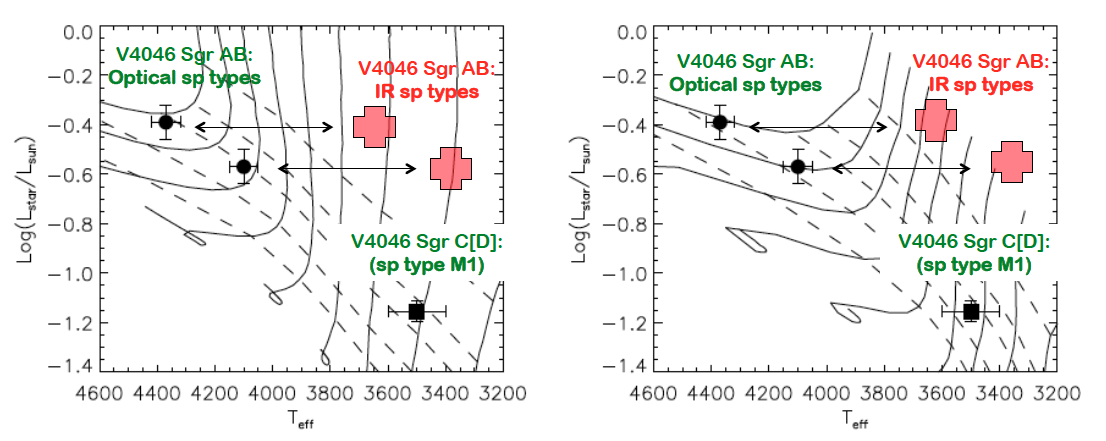}
\caption{H-R diagram positions of V4046 Sgr AB and its wide-separation M1 companion (V4046 Sgr C[D]) as inferred optically (thin black crosses) and from IRTF/SpeX data (\S 3; thick red crosses) overlaid on pre-MS
evolutionary tracks from ({\it left}) \citet{2000A&A...358..593S} and ({\it right}) \citet{1998A&A...337..403B}. Masses (solid lines) run from 0.2 to 1.0 $M_\odot$ in intervals
of 0.1 $M_\odot$, and the isochrones (dashed lines) are for ages 5, 10, 15, 25, and 40 Myr. Adapted from \citet{2011ApJ...740L..17K}.
}
\label{fig:HRdiagrams}
\end{center}
\end{figure}

The 3--5 spectral subclass discrepancy we have found between optical and near-IR (IRTF/SpeX-determined) spectral types for V4046 Sgr AB is very similar to that found for TW Hya by \citet{2011ApJ...732....8V}. However, in contrast to the case of the (nearly pole-on, single) TW Hya star/disk system, the component masses of the V4046 Sgr AB (binary plus circumbinary disk) system are very well constrained, thanks to its intermediate system inclination \citep[$33.5^\circ$;][]{2012ApJ...759..119R}. 
%Specifically, comparison of the central mass and luminosity of V4046 Sgr with pre-MS tracks imposes a combination of component masses in the ranges 0.85--0.95 $M_\odot$ (A) and 0.80--0.90 $M_\odot$ (B) and age in the range 6--30 Myr \citep{2012ApJ...759..119R}.  
Given these constraints, the optically determined spectral type combination of K5+K7 and, hence, ``mean'' corresponding effective temperature of $T_{eff}$ $\sim$4300 K \citep{2011MNRAS.417.1747D} is in much better agreement with the predictions of pre-MS evolutionary tracks, given the likely system age range (12--21 Myr), than is the ``mean'' of $T_{eff}\sim3500$ K that corresponds to the early-M spectral type implied by our analysis of IRTF/SpeX spectroscopy (Fig.~\ref{fig:HRdiagrams}). 

These results further imply that the SPEX-based inference of an M2.5 spectral type for TW Hya \citep{2011ApJ...732....8V} should not be used to conclude that this heavily scrutinized pre-MS star is younger and less massive than implied by its optical spectral type \citep[see also discussions of the mass and age of TW Hya in][]{2012ApJ...744..162A,2014A&A...563A.121D}.
%Indeed, the latter (late-K) spectral type is in better agreement with the predictions of pre-MS tracks for the generally accepted stellar mass of $\sim$0.8 Msun determined from CO kinematics \citep{2012ApJ...744..162A}, given the age of the TW Hya Association \citep[$\sim$8 Myr;][and refs.\ therein]{2014A&A...563A.121D}. 
We note that \citet{2013ApJ...771...45D} found good fits either for a visual/near-IR ``composite'' of K7+M2 or for a ``compromise'' spectral type of M0 to an HST/STIS spectrum of TW Hya that featured very broad spectral coverage (extending from $\sim$5000--10000 \AA). They infer a mass as small as 0.55 $M_\odot$, if the underlying star is better represented by the cooler (M2) component, which would be the case if the composite spectral type can be ascribed to the effects of accretion (see below). However, our IRTF/SpeX results for V4046 Sgr suggest that an earlier spectral type, hence higher mass, is preferred for TW Hya as well. 

The likely causes of the systematic discrepancies between optical and near-IR spectral types are starspots and surface gravity effects, as discussed by \citet{1998ASPC..154.1709G} and \citet{2013ApJ...769...73M}. \citet{2013ApJ...771...45D} propose an additional explanation, in which accretion {\it hot} spots make a cooler photosphere appear warmer as one moves blueward in the optical from the near-IR. However, the nonaccreting (weak-lined) TTS in Taurus studied by \citet{1998ASPC..154.1709G} also appear too cool in the near-IR relative to MS stars of the same optically-determined spectral types, casting doubt on the importance of accretion in producing the spectral type discrepancies for the \citep[weakly accreting;][]{2011A&A...526A.104C} TW Hya and V4046 Sgr AB. An important test case in this regard is V4046 Sgr C[D], the comoving companion to AB \citep{2011ApJ...740L..17K}; V4046 Sgr C[D] is a close-binary, weak-lined T Tauri system with a composite optical spectral type of M1e. Given that AB and C[D] are presumably coeval and their relative luminosities are well-established (Fig.~\ref{fig:HRdiagrams}), comparison of the results of near-infrared spectroscopy of the (accreting) V4046 Sgr AB and (nonaccreting) C[D] systems should provide stringent constraints on the physical origins of optical vs.\ near-IR spectral type discrepancies for pre-MS stars.

\section{Conclusions}

We have shown that the close binary pre-MS star system V4046 Sgr AB exhibits a (composite binary) spectral type of early M in the near-IR, i.e., significantly (3--5 subtypes) later than the mid/late-K spectral types previously determined via analysis of its optical spectrum. This discrepancy is very similar to that displayed by the single pre-MS star TW Hya \citep{2011ApJ...732....8V}. It is likely that the combination of large surface area starspots and low surface gravities, both of which are characteristic of pre-MS stars, results in a trend of later spectral type with increasing wavelength, moving from the optical to the near-infrared regime. 

Regardless, it is apparent --- given the well-constrained mass and age of the V4046 Sgr binary --- that the optical spectral type determination for V4046 Sgr AB is in better agreement with the predictions of evolutionary models than our near-infrared spectral type determination. {\it These results suggest that the use of near-infrared spectra as a primary means to determine pre-MS star spectral types may lead to significant underestimates of photospheric temperatures, which can in turn produce significant, systematic errors in inferred pre-MS star masses and ages.}

Methods similar to those employed here for V4046 Sgr and by \citet{2011ApJ...732....8V} for TW Hya should be applied to additional pre-MS stars, both accreting and non-accreting, whose masses can be determined via independent means \citep[e.g., via CO kinematics or, for the growing sample of known eclipsing binary T Tauri systems, via binary orbital parameters;][]{2014ATsir1610....1I}. Given such near-IR spectral type determinations for a large and diverse sample of pre-MS stars with precisely-determined masses and well-established ages and luminosities, we can more fully characterize and calibrate optical vs.\ near-IR spectral type discrepancies for pre-MS stars --- which, based on the two cases considered here, would appear to be $\sim$3--5 spectral subclasses for weakly accreting, roughly solar-mass pre-MS stars with ages in the range $\sim$10--20 Myr. 

\acknowledgments{Support for this research is provided by by National Science Foundation grant AST--1108950 to RIT. C.T. Smith's research at RIT was supported by a NSF Research Experience for Undergraduates program grant to RIT's Chester F. Carlson Center for Imaging Science.}


\begin{thebibliography}{1}

\bibitem[Andrews et al.(2012)]{2012ApJ...744..162A} Andrews, S.~M., Wilner, 
D.~J., Hughes, A.~M., et al.\ 2012, \apj, 744, 162 

\bibitem[Baraffe et 
al.(1998)]{1998A&A...337..403B} Baraffe, I., Chabrier, G., Allard, F., \& Hauschildt, P.~H.\ 1998, \aap, 337, 403 


\bibitem[Binks 
\& Jeffries(2014)]{2014MNRAS.438L..11B} Binks, A.~S., \& Jeffries, R.~D.\ 2014, \mnras, 438, L11 

\bibitem[Curran et 
al.(2011)]{2011A&A...526A.104C} Curran, R.~L., Argiroffi, C., Sacco, G.~G., et al.\ 2011, \aap, 526, A104 

\bibitem[Cushing et al.(2004)]{2004PASP..116..362C} Cushing, M.~C., Vacca, 
W.~D., \& Rayner, J.~T.\ 2004, \pasp, 116, 362 

\bibitem[Debes et al.(2013)]{2013ApJ...771...45D} Debes, J.~H., 
Jang-Condell, H., Weinberger, A.~J., Roberge, A., 
\& Schneider, G.\ 2013, \apj, 771, 45 


\bibitem[Donati et al.(2011)]{2011MNRAS.417.1747D} Donati, J.-F., Gregory, 
S.~G., Montmerle, T., et al.\ 2011, \mnras, 417, 1747 

\bibitem[Ducourant et 
al.(2014)]{2014A&A...563A.121D} Ducourant, C., Teixeira, R., Galli, P.~A.~B., et al.\ 2014, \aap, 563, A121 


\bibitem[Gullbring et al.(1998)]{1998ASPC..154.1709G} Gullbring, E., 
Hartmann, L., Briceno, C., Calvet, N., 
\& Muzerolle, J.\ 1998, Cool Stars, Stellar Systems, and the Sun, 154, 1709 

\bibitem[Herbig(1978)]{herbig78} Herbig, G. H. 1978, in Problems of Physics and Evolution of the Universe, ed.
L. V. Mirzoyan (Yerevan: Publ. Armenian Academy of Science), 171

\bibitem[Ismailov et al.(2014)]{2014ATsir1610....1I} Ismailov, N.~Z., Abdi, 
H.~A., \& Mamedxanova, G.~B.\ 2014, Astronomicheskij Tsirkulyar, 1610, 1 


\bibitem[Kastner et al.(2011)]{2011ApJ...740L..17K} Kastner, J.~H., Sacco, 
G.~G., Montez, R., et al.\ 2011, \apjl, 740, L17 

\bibitem[Kastner(2013)]{kastner13} Kastner, J.~H. 2013, in ``The Star Formation Newsletter,'' ed. B. Reipurth, No.\ 245 (\verb+http://www.ifa.hawaii.edu/~reipurth/newsletter/newsletter245.pdf+)

\bibitem[Kastner et al.(2014)]{kastner14} Kastner, J.~H., Hily-Blant, P., Rodriguez, D., Punzi, K., \& Forveille, T. 2014, \apj, in press (arXiv: 1408.5918)

\bibitem[Mann et al.(2013)]{2013AJ....145...52M} Mann, A.~W., Brewer, 
J.~M., Gaidos, E., L{\'e}pine, S., \& Hilton, E.~J.\ 2013, \aj, 145, 52 

\bibitem[McClure et al.(2013)]{2013ApJ...769...73M} McClure, M.~K., Calvet, 
N., Espaillat, C., et al.\ 2013, \apj, 769, 73 

\bibitem[Pecaut 
\& Mamajek(2013)]{2013ApJS..208....9P} Pecaut, M.~J., \& Mamajek, E.~E.\ 2013, \apjs, 208, 9 

\bibitem[Rayner et al.(2003)]{2003PASP..115..362R} Rayner, J.~T., Toomey, 
D.~W., Onaka, P.~M., et al.\ 2003, \pasp, 115, 362 

\bibitem[Rayner et al.(2009)]{2009ApJS..185..289R} Rayner, J.~T., Cushing, 
M.~C., \& Vacca, W.~D.\ 2009, \apjs, 185, 289 

\bibitem[Rosenfeld et al.(2012)]{2012ApJ...759..119R} Rosenfeld, K.~A., 
Andrews, S.~M., Wilner, D.~J., \& Stempels, H.~C.\ 2012, \apj, 759, 119 

%\bibitem[Rosenfeld et al.(2013)]{2013ApJ...775..136R} Rosenfeld, K.~A., 
%Andrews, S.~M., Wilner, D.~J., Kastner, J.~H., \& McClure, M.~K.\ 2013, \apj, 775, 136 
\bibitem[Siess et 
al.(2000)]{2000A&A...358..593S} Siess, L., Dufour, E., \& Forestini, M.\ 2000, \aap, 358, 593 

\bibitem[Stempels 
\& Gahm(2004)]{2004A&A...421.1159S} Stempels, H.~C., \& Gahm, G.~F.\ 2004, \aap, 421, 1159 

\bibitem[Torres et 
al.(2006)]{2006A&A...460..695T} Torres, C.~A.~O., Quast, G.~R., da Silva, L., et al.\ 2006, \aap, 460, 695 

\bibitem[Torres et al.(2008)]{2008hsf2.book..757T} Torres, C.~A.~O., Quast, 
G.~R., Melo, C.~H.~F., 
\& Sterzik, M.~F.\ 2008, Handbook of Star Forming Regions, Volume II, 757 

\bibitem[Vacca et al.(2003)]{2003PASP..115..389V} Vacca, W.~D., Cushing, 
M.~C., \& Rayner, J.~T.\ 2003, \pasp, 115, 389 

\bibitem[Vacca 
\& Sandell(2011)]{2011ApJ...732....8V} Vacca, W.~D., \& Sandell, G.\ 2011, \apj, 732, 8 

\end{thebibliography}
\end{document}